\documentclass[pre,twocolumn,showpacs,preprintnumbers,amsmath,amssymb]{revtex4}
\usepackage{graphicx}
\usepackage{dcolumn}
\usepackage{bm}

\begin{document}

\title{Giant slip lengths of a simple fluid at vibrating solid interfaces}
\author{Aur\'elien Drezet $^1$, Alessandro Siria$^{1,2}$, Serge Huant$^1$ and Jo\"{e}l Chevrier$^{1}$}
\affiliation{$^{1}$ Institut N\'eel, CNRS and Universit\'e Joseph Fourier Grenoble, BP 166 38042 Grenoble Cedex 9, France\\
$^{2}$ CEA/LETI-MINATEC, 17 Avenue des Martyrs 38054 Grenoble Cedex 9, France.}

\begin{abstract}
It has been shown recently [PRL \textbf{102}, 254503 (2009)] that
in the plane-plane configuration a mechanical resonator vibrating close to a rigid wall in a simple
fluid can be overdamped to a frozen regime. Here, by solving analytically the Navier Stokes equations with
partial slip boundary conditions at the solid fluid interface, we
develop a theoretical approach justifying and extending these
earlier findings. We show in particular that in the perfect slip regime the above mentioned results are, in the plane-plane configuration, very general and robust with respect to lever geometry considerations. We compare the results with those obtained previously for the sphere moving perpendicularly and close to a plane in a simple fluid and discuss in more details the differences concerning  the dependence of the friction forces with the gap distance separating the moving object (i.e., plane or sphere) from the fixed plane. Finally, we show that the submicron fluidic
effect reported in the reference above, and discussed further in the present work, can have dramatic implications in the design of
nano-electromechanical systems (NEMS).
\end{abstract}
 \maketitle
\section{Introduction}
Nanomechanical resonators and nanoelectromechanical systems are
widely used in a multitude of applications such as ultrafast
actuation and sensing at the zeptogram and sub-attonewton
scales~\cite{Roukes2007,cooling,Ekinci,Cap1,Cap3}. The
extraordinary sensitivity that such NEMS provide rely mainly upon
their very high oscillating pulsations and quality factors $Q$.
However, while  $Q$ factors in the tens or hundreds of thousands
can be obtained in vacuum and/or cryogenic environment, these
values degrade dramatically in liquid and gas phases meaning that
much more work is still to be done to reach the technological
level (see however~\cite{Ekinci, rugar01}). It is therefore
necessary to characterize more precisely the viscous forces
exerted by fluids on the vibrating motion of
NEMS and this constitutes the motivation for the present work.\\
Over the last years, micro- and nano-fluidics experiments
\cite{Tabeling} reported that the physical properties of fluids
flowing into or around confined systems, such as
nano-channels~\cite{nanoc}, are strongly modified compared with
those encountered at the micro and macro scales. In particular, it
has been shown that the usual no-slip boundary conditions, which
are universally used since the $19^{th}$ century to model the
behavior of Newtonian fluids at a solid interface, break down at
the nano scale~\cite{Lauga}. Such modifications of boundary
conditions are also expected to have a huge impact on NEMS
dynamics since properties known for Microelectromechanical systems
(MEMS)~\cite{Green,Naik,Paul,Dorignac,Basak,Tung} cannot simply be
scaled down to the nano realm.\\ In a recent work, we investigated
the importance of gas damping on a thermally actuated microlever
as it is gradually  approached towards an infinite wall in
parallel geometry~\cite{nous}. The experiment performed at room
temperature in air showed that the sub-Angstrom lever oscillation
amplitude, i.e., in the direction perpendicular to the parallel
planes, is completely frozen as the gap $d$ is progressively
decreased from 20 $\mu$m to 400 nm. Moreover, the friction force
recorded was much larger that predicted by the Navier-Stokes
hydrodynamical equations solved together with the no-slip boundary
conditions. Instead, the reported results are qualitatively and
quantitatively well understood if one accepts the perfect-slip
boundary conditions for which friction at the lever/gas interface
is prohibited.\\
The aim of the present article is twofold. First,
we study theoretically the motion of the Newtonian fluid, i.e.
air, around the oscillating micro plate. Starting with the
linearized Navier-Stokes equation we analytically show that
perfect slip boundary conditions lead to the correct dynamical
behavior reported experimentally in ref.~\cite{nous}. We compare
our findings with the other more traditional approaches based on
the no-slip boundary conditions and show that they necessarily
conflict with the experimental facts. During the analysis we also
discuss the different possible boundary conditions and in
particular the impact of the slip length of the beam dynamical
behavior. The second goal of this paper is to show the important
implications that our findings may have on the engineering and
architecture of future NEMS operating in gaseous environment.
Here, on the basis of the results obtained in ref.~\cite{nous} for
a model system we discuss precisely the existence of a critical
overdamped regime for NEMS oscillating in fluids and study the
influence of materials, geometrical, and intrinsic dynamical
properties on this regime.\\
The paper is organized as follows. In
Sec.~II we discuss the general characteristics of the dynamic
problem in conjunction with Navier-Stokes equations and boundary
conditions. In Sec.~III the static regime corresponding to various
slip lengths and valid for small gap values is studied
analytically and compared (successfully) with the experimental
results reported in ref.~\cite{nous}. In sec.~IV we briefly discuss the possible microscopic mechanisms involved in order to explain the reported results. In particular, we compare the calculations obtained here for the plane-plane geometry to those already obtained in the sphere-plane configuration by Taylor and Vinogradova~\cite{Vinogradova,Vinogradova2}.  Finally, in Sec.~V we
discuss the consequences of our findings for NEMS dynamics.  A
summary is given in Sec.~VI.
\section{Mechanical oscillator in a newtonian fluid}

We consider a thin silicon commercial cantilever used in atomic
force microscopy (AFM) for liquid imaging~\cite{nous}. This system
is modelled as a parallelepiped with length $L$, width $w$ and
thickness $t$. The lever is clamped by one end and oscillates
mainly along the $z$ direction (see Fig.~1).
\begin{figure}[hbtp]
\includegraphics[width=0.7\columnwidth]{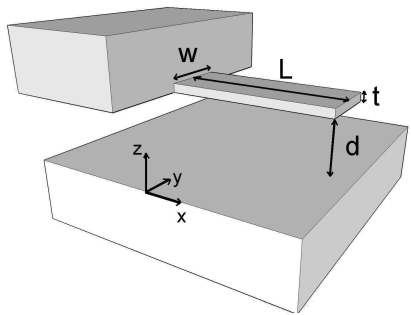}\\
\caption{Scheme of the one end-clamped parallelepiped cantilever
used in ref.~\cite{nous}. The lever has dimensions $t$, $L$, and
$w$ and is vibrating in the $z$ direction at a distance $d$ from
the substrate. }
\end{figure}
In vacuum this beam may be viewed as a harmonic oscillator with an
intrinsic resonance pulsation $\omega_0$ and effective mass $m$
(internal losses can be fairly neglected in the following
analysis). In the fluid, the viscous force acting on the lever is
characterized by a dissipative coefficient $\gamma$ connecting the
viscous force $F_z$ normal to the lever to the velocity $U$ of the
lever along the same direction: $F_z=-\gamma\cdot U$. At short
distance, i.e, in the non retarded regime, $\gamma$ is given
by~\cite{nous}:
\begin{equation}
 \gamma= \frac{2 \eta  Lw}{d},
\end{equation}
where $\eta$ is the (dynamic) fluid viscosity and $d$ the distance
between the lever and the substrate (see Fig.~1). As we will now
show this law results directly from the perfect slip boundary
conditions applied to the Navier-Stokes equation, and it differs
from the usual $\gamma\simeq \eta w L^3/d^3$ behavior deduced by
considering the
no slip conditions~\cite{nous,landau}.\\
To derive this result we start from  the non-linear Navier-Stokes equations for an incompressible fluid
\begin{equation}
\rho \left[ \partial_t \vec v+(\vec v \cdot\vec{\nabla}) \vec v\right] = \eta \vec{\nabla} ^2 \vec v - \vec{\nabla} p \qquad
\label{eq:ns}
\end{equation}
where $\vec v$ is the local fluid velocity, $\rho$ its density,
and $p$ the gas pressure. To calculate the gas flow around the
lever we take into account specifical properties of the system
under study. First, the importance of the non-linear term $(\vec v
\cdot\vec{\nabla}) \vec v$ can be estimated from the knowledge of
the Reynolds number $R_e:=v \cdot X/\nu=\rho v\cdot X/\eta$
calculated for a characteristic length $X$ and fluid velocity $v$
($\nu=\eta/\rho=1.5\times 10^{-5}$ m$^{2}$/s is the kinematic
viscosity). In the present context, a correct order of magnitude
of the velocity in the fluid is given by the lever velocity
components $U_x$, $U_z:=U$ along the x and z direction
respectively. An important related feature is  that we have here
$U_x\ll U$. Indeed, writing $\theta \simeq \delta z/L$ (with
$\delta z=0.05$ nm the typical lever oscillation amplitude) the
main angle shown by the lever with the $x$ axis we get the
estimation $U_x/U\simeq \theta \simeq 10^{-6} $ which implies that
the motion is mainly vertical. To simplify our analysis we
therefore assume that the lever is a horizontal plate vibrating
along the z direction.  The lever oscillating at the frequency
$\omega/(2\pi)=50$ $k$Hz one obtains $U\simeq\omega \delta
z/(2\pi)\simeq 2.5 \times10^{-6}$ m/s and thus with $X:=d=50$
$\mu$m
\begin{eqnarray}
R_e=\frac{\omega \delta  z \cdot d}{2\pi\nu}\simeq 0.8\times 10^{-5}\ll 1.
\end{eqnarray}
As a consequence of this ultra-small Reynolds number we will
completely neglect nonlinearity in the rest of this work. As a
corollary of this analysis we also deduce the Mach number
$M=v/c\sim U/c\simeq 10^{-8}\ll 1$ ($c$ sound velocity in air).
The vanishing value of $M$ justifies the fluid incompressibility hypothesis $\vec{\nabla}\cdot \vec v=0$.\\
The second question that we should deal with concerns the
amplitude of the dynamical term $\rho \partial_t \vec
v=-i\omega_0\rho\vec v$ in Eq.~\ref{eq:ns}. Since we are concerned
with harmonic oscillations we introduce a second Reynolds number:
\begin{eqnarray}
R'_e=\frac{\omega_0 \cdot d^2}{\nu}\simeq 2\times 10^{-2}\cdot
(d[\mu m])^2,
\end{eqnarray}
where $d[\mu m]$ is the measure of $d$ in micrometers. Clearly  $R'_e$ (and therefore $\rho \partial_t \vec v$) is negligible as far as $d \ll\sqrt{2\nu/\omega_0}\simeq 10$ $\mu$m, i.e. as far as the gap $d$ is smaller  than the boundary layer thickness $\delta_B$ (see below for discussion). In the present work we will only consider this static regime and neglect the dynamic term.\\
In order to solve the linearized Navier-Stokes equation
\begin{equation}
\label{eq:nsL} \eta \vec{\nabla} ^2
\vec v - \vec{\nabla} p\simeq 0,
\end{equation}
one must provide the precise boundary conditions for the fluid velocity at the solid interfaces. The condition
for the normal component of the velocity is quite obvious~\cite{landau,batchelor}. Indeed, since the fluid cannot
go through a solid interface the fluid velocity  $v_z$  must equal the velocity of the plate $U$ at $z=d$ and must
also vanish along the surface $z=0$~\cite{landau,batchelor}. However, the precise form of the conditions for the tangential
components $v_x$ and $v_y$ is not so natural and is a subject of debates and controversies since the birth
of hydrodynamics~\cite{Lauga,Loeb,Neto,Charlaix}.  The problem was already well addressed by Navier~\cite{Navier} who introduced the two most known
possibilities which are respectively the no-slip and perfect slip boundary conditions. Following
the no-slip hypothesis the fluid velocity along the interface must equal the in-plane velocity of the solid boundary. In the present case the no-slip conditions lead to $v_x=v_y=0$ along the substrate plane $z=0$ and  $v_x=U_x\simeq0$, and $v_y=0$ along the cantilever interface at $z=d$. The no-slip hypothesis is well documented in the literature and experimentally justified at the macroscale~\cite{Lauga,Loeb,Neto,Charlaix}. It leads however to increasing difficulties and contradictions in the micro and nanofluidic regime where fundamental and statistical properties of molecules such as the mean free path and the surface roughness cannot be ignored~\cite{Lauga,Loeb,Neto,Charlaix,Tabeling}.\\
The second extreme possibility, the perfect slip boundary
conditions, suppose having $\partial_zv_i=0$ (from now on we use
the notation $i=x,y$) along the solid interface. These are
reminiscent for conditions on the viscous stress tensor
$\sigma_{iz}=\eta(\partial_zv_i+\partial_iv_z)=0$ meaning that no
tangential friction is allowed between fluid and solid. We however
point out that the equivalence between the relations
$\sigma_{iz}=0$ and $\partial_zv_i=0$ assume the additional
conditions $\partial_iv_z=0$ which implies that $v_z$ does not
depend on  $x,y$ in the vicinity of the solid boundary (this is
obviously true on the interface provided that the boundary
conditions on $v_z$ are fulfilled). Beyond these two extreme cases
more realistic approaches were proposed to take into account a
possible partial slip at the boundary. In particular
Navier~\cite{Lauga,Loeb,Neto,Charlaix,Tabeling,Navier} suggested
the existence of a surface friction force law
$\sigma_{iz}n_z=\kappa v_i$ where $\vec n=n_z \hat{z}$ is a unit
vector normal to the surface and oriented outwardly from the solid
to the fluid and $\kappa$ a dissipative coefficient. With the same
assumptions as before this leads to
\begin{equation}
 n_z\partial_zv_i=\kappa v_i/\eta=v_i/b,\label{slipl}
\end{equation}
where $b$ is the so called slip length. This law has been
considerably studied in the recent years with the advent of micro
and nanofluidics~\cite{Lauga,Neto,Charlaix,Tabeling}. Eq.~\ref{slipl}  has been confirmed many times
in particular for liquid flows in nanochannels~\cite{Neto}. However, the
measurements of the associated slip length $b$ reveals a broad
spectrum of values which specifically depend on the system
considered~\cite{Tabeling,Charlaix}. This shows that only a microscopic analysis could lead
to a better understanding of the phenomenon. Keeping this point
for latter discussions we will here apply the law given by
Eq.~\ref{slipl} to our problem and see  how it compares  with the
experimental results.
\section{The static limit }
In the present analysis we consider the static regime valid for $d
\ll \delta_B:=\sqrt{2\nu/\omega_0}$. We must therefore solve the
system of coupled equations $\eta \vec{\nabla} ^2 \vec v =
\vec{\nabla} p $, $\vec{\nabla}\cdot \vec v=0$ together with the
conditions given by Eq.~\ref{slipl}. The problem is
reminiscent for the one solved by Reynolds in the case of two
parallel disks in a dissipative fluid. Reynolds considered two
disks approaching each other with a constant velocity $\pm U$
along their common axis of symmetry. However, despite geometry
differences, Reynolds considered specifically the case of no-slip
boundary conditions (which were universally accepted at that time)
and not the more general Eq.~\ref{slipl}. Using the same
approximations than Reynolds we here suppose $\partial_x
v_i,\partial_y v_i\ll\partial_z v_i$, $\partial_x v_z,\partial_y
v_z\ll\partial_z v_z$, and  $\partial_z p\simeq0$ which are
standard in lubrication theory. We then have
\begin{eqnarray}
\eta\partial^2_zv_i\simeq\partial_i p.
\end{eqnarray}
After integration with respect to $z$ this leads to
\begin{equation}
v_i(x,y,z)=\frac{1}{2\eta}\partial_i
p(x,y)z^2+\alpha_i(x,y)z+\beta_i(x,y).
\end{equation}
To be general we are introducing two \emph{a priori} different
slip lengths $b_0$ and $b_1$ for the interfaces at $z=0$ and $z=d$,
respectively. This hypothesis implies
\begin{eqnarray}
v_i(z=0)=b_0\partial_z v_i(z=0), & v_i(z=d)=-b_1\partial_z v_i(z=d)\nonumber,
\end{eqnarray} and therefore
\begin{equation}
v_i(x,y,z)=\frac{1}{2\eta}\partial_i
p(x,y)[z^2-d\frac{2b_1+d}{b_0+b_1+d}(z+b_0)].
\end{equation}
The fluid incompressibility relation $\partial_x v_x+\partial_y
v_y+\partial_zv_z=0$ and the boundary conditions for $v_z$ at
$z=0$ give the expression: $v_z(x,y,z)=-\int_0^z(\partial_x
v_x+\partial_y v_y) dz$, i.e.,
\begin{eqnarray}
v_z=\frac{1}{2\eta}(\partial_x^2+\partial_y^2)p(x,y)[d\frac{2b_1+d}{b_0+b_1+d}(\frac{z^2}{2}+b_0z)-\frac{z^3}{3}].\nonumber\\
\end{eqnarray}
At $z=d$ we have the boundary condition $v_z=U$ and we deduce
\begin{eqnarray}
(\partial_x^2+\partial_y^2)p(x,y)=\frac{2\eta U}{[(-\frac{1}{3}+\frac{1}{2}\frac{2b_1+d}{b_0+b_1+d})d^3+\frac{2b_1+d}{b_0+b_1+d}b_0 d^2]}\nonumber\\
\end{eqnarray}
and
\begin{eqnarray}
v_z(x,y,z)=\frac{U[-\frac{z^3}{3}+d\frac{2b_1+d}{b_0+b_1+d}(\frac{z^2}{2}+b_0z)]}{[(-\frac{1}{3}+\frac{1}{2}\frac{2b_1+d}{b_0+b_1+d})d^3+\frac{2b_1+d}{b_0+b_1+d}b_0
d^2]}. \label{eq:vzc}
\end{eqnarray}
In particular, in  the limit $b_0=b_1\rightarrow +\infty$ we have
$v_z=Uz/d$ whereas for $b_0=b_1=0$ we obtain
$v_z=6U(-z^3/3+z^2d/2)/d^3$. It should be observed that the
solution for $v_z$ in the perfect slip limit looks like the well
known solution of the Couette problem ~\cite{landau,batchelor} for
the permanent fluid motion between two plates in relative and
uniform  motion along the $x$ direction. It is worth noting that
the direction of the fluid motion and the boundary conditions used
are however completely different in these two problems (indeed, in
the Couette problem we assume the no-slip boundary conditions and
neglect $v_z$~\cite{landau,batchelor}).
Here due to the presence of terms with $v_z$ our approach deviates from this standard result.\\
The normal force exerted by the fluid located between $z=0$ and
$z=d$ on each surface element $dxdy$ of the lever is given by
\begin{equation}
dF_z=[-\sigma_{zz}+p]|_{z=d}dxdy=[-2\eta\partial_zv_z|_{z=d}+p(x)]dxdy.
\end{equation}
$\sigma_{zz}=2\eta\partial_zv_z$ is the fluid stress tensor along
the $z$ direction and therefore $-\sigma_{zz}|_{z=d}dxdy$ is the
dissipative part of the resulting force $dF_z$. Using
Eq.~\ref{eq:vzc} we obtain
\begin{eqnarray}
-\sigma_{zz}|_{z=d}=\frac{2\eta
U[d^2-\frac{2b_1+d}{b_0+b_1+d}(d^2+b_0
d)]}{[(-\frac{1}{3}+\frac{1}{2}\frac{2b_1+d}{b_0+b_1+d})d^3+\frac{2b_1+d}{b_0+b_1+d}b_0
d^2]}.
\end{eqnarray}
Interestingly this contribution is independent of $x,y$ and of the
lateral boundary shape $(C)$ associated with the lever and the
substrate.  The resulting dissipative force
$-\int_{(S)}\sigma_{zz}|_{z=d}dxdy$ on the plate of surface $S$ is
consequently
\begin{eqnarray}
F_z^{\textrm{dissip.}}=\frac{2\eta
SU[d^2-\frac{2b_1+d}{b_0+b_1+d}(d^2+b_0
d)]}{[(-\frac{1}{3}+\frac{1}{2}\frac{2b_1+d}{b_0+b_1+d})d^3+\frac{2b_1+d}{b_0+b_1+d}b_0
d^2]}.
\end{eqnarray}
The second contribution to the force is associated with the volume
pressure $p(x,y)$ that is solution of the 2D Poisson equation
$(\partial_x^2+\partial_y^2)p(x,y)=A$ where $A$ is a constant (see
Eq.~11). Lateral boundary conditions along $(C)$ are here playing
explicitly a role in the analysis. Let us suppose for example that
the substrate is infinitely extended in the $z=0$ plane and that
the lever is delimited by a rectangular boundary of length $L$ and
width $w$ in the $z=d$ plane. A lengthy Fourier analysis (see
appendix) shows that the pressure field can be expanded as:
\begin{eqnarray}
p(x,y)=p_0+\frac{4AL^2}{\pi^3}\sum_n\frac{1}{n^3}\sin{(\pi
nx/L)}[\alpha_n e^{+\pi ny/L}\nonumber\\ +\beta_n e^{-\pi n
y/L}-1],
\end{eqnarray}
where $n$ are odd integers and $\alpha_n,\beta_n$ are constant
coefficients (see appendix). To obtain this formula the boundary
condition $p=p_0$ ($p_0$ is the atmospheric pressure) was imposed
on the rectangular contour. Similarly, if we consider the 1D
problem with $L$ finite, $w=+\infty$, and with $p$ a function of
$x$ only we deduce after  a direct integration
\begin{equation}
p(x)=p_0+\frac{A}{2}(x^2-xL),\label{truc}
\end{equation} where we used the boundary conditions $p(0)=p(L)=p_0$. Equivalently, this
result could also be obtained from the Fourier expansion
$p(x)=p_0-\frac{4AL^2}{\pi^3}\sum_n\frac{1}{n^3}\sin{(\pi nx/L)}$.
Finally, as a last example, we consider the circular plate of
radius $R$. Assuming the radial symmetry this problem leads to the
solution
\begin{equation}
p(r)=p_0+\frac{A}{4}(r^2-R^2),\label{retruc}
\end{equation}
where $r$ is the radial coordinate and $p(R)=p_0$ along the
circular contour of radius $R$. More generally, by using the
uniqueness theorem it is easily shown that the pressure  $p(x,y)$
can always be written
\begin{equation}
p(x,y)=p_0+A\cdot g(x,y),
\end{equation}
where $g(x,y)$ is the solution of
$(\partial_x^2+\partial_y^2)g(x,y)=1$  which is univocally
determined  by the boundary condition $g(x,y)=0$ along the
contour. The resulting vertical force due to the pressure field on
the plate of surface $S$ is thus given by
\begin{eqnarray}
F_z^{\textrm{pressure}}=\int\int_{(S)} (p(x,y)-p_0)dxdy\nonumber\\=A\int\int_{(S)} g(x,y)dxdy=AS\langle g\rangle,
\end{eqnarray}
where the term $-p_0$  equilibrating the pressure $+p_0$ from
Eq.~19 is due to the force exerted by the fluid on the second side
of the cantilever.  Considering the examples quoted previously we
find explicitly
\begin{eqnarray}
F_z^{\textrm{pressure}}=-\frac{8AL^3w}{\pi^4}\sum_n\frac{1}{n^4}[1\nonumber\\-\frac{2L}{\pi
nw}\frac{(e^{+\pi n w/(2L)}-e^{-\pi n w/(2L)})^2}{e^{+\pi n
w/L}-e^{-\pi n w/L}}],
\end{eqnarray}
for the rectangular plate, and
\begin{eqnarray}
F_z^{\textrm{pressure}}=-\frac{8AL^3w}{\pi^4}\sum_n\frac{1}{n^4}=-\frac{AL^3w}{12},
\end{eqnarray}
for the 1D plate (note that $w$ is a finite width in the $y$
direction introduced for reasons of dimensionality and that Eq.~22
is the limit of Eq.~21 for $w\rightarrow+\infty$). Similarly, for
the disk we deduce
\begin{eqnarray}
F_z^{\textrm{pressure}}=-\frac{\pi AR^4}{8}.
\end{eqnarray}
These forces due to pressure can be compared with the dissipative
contribution given by Eq.~15 which can equivalently be expressed
as
\begin{eqnarray}
F_z^{\textrm{dissip.}}=AS[d^2-\frac{2b_1+d}{b_0+b_1+d}(d^2+b_0 d)],
\end{eqnarray}
and leads to the total force $F_z=F_z^{\textrm{pressure}}+F_z^{\textrm{dissip.}}$:
\begin{eqnarray}
F_z=AS[\langle g\rangle+d^2-\frac{2b_1+d}{b_0+b_1+d}(d^2+b_0 d)],
\end{eqnarray}
with
\begin{eqnarray}
A=\frac{2\eta
U}{[(-\frac{1}{3}+\frac{1}{2}\frac{2b_1+d}{b_0+b_1+d})d^3+\frac{2b_1+d}{b_0+b_1+d}b_0
d^2]}.
\end{eqnarray}
In the usual limit of vanishing slip lengths
$F_z^{\textrm{dissip.}}=0$ and we therefore have
\begin{eqnarray}
F_z=\frac{12\eta US\langle g\rangle}{d^3},
\end{eqnarray} which is the generalization of the Reynolds formula for an arbitrary plate shape in the no-slip limit.
Inversely, in the limit $b_0=b_1$ infinite, i.e, perfect slip, we obtain $p(x,y)=p_0$, $F_z^{\textrm{pressure}}=0$ and thus
\begin{eqnarray}
F_z=-\frac{2\eta US}{d},
\end{eqnarray} which is the result used in ref.~\cite{nous} in the perfect slip
limit. It is worth noting that this formula is only dependent on
the surface $S$  of the vibrating plate and not on its exact
geometry. It implies that the result obtained in ref.~\cite{nous}
should be very robust with respect to geometry considerations.
\begin{figure}[hbtp]
\includegraphics[width=8.5cm]{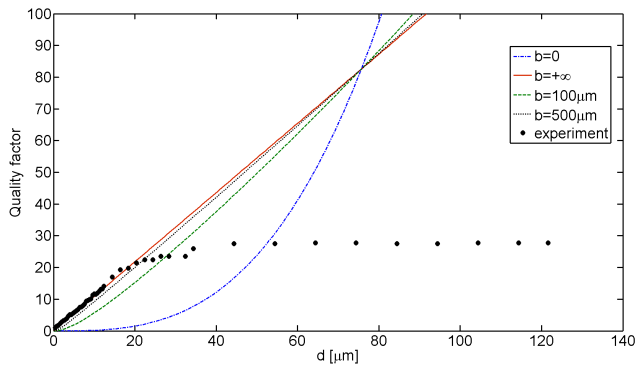}\\
\caption{Evolution of the beam quality factor $Q$ as a function of the gap $d$ for different slip lengths $b$ and for the lever mechanical properties given in ref.~\cite{nous}.  The experimental data are in good agreement only with  $b> 500$ $\mu m$ (red curve).  }
\end{figure}
We point out that this perfect-slip limit can be directly obtained by
solving the equation $\eta\partial^2_zv_z=0$ with the boundary
conditions for $v_z$. Indeed, the direct integration gives $v_z=U
z/d$. The pressure $p=p_0$ can be deduced
similarly~\cite{remarque}.\\
It is worth noting
that in the derivation of all these formulas the Reynolds
assumption $\partial_zp=0$ implies also
$\eta\partial^2_zv_z=\partial_z p=0$. However, from Eq.~12 we
deduce
\begin{eqnarray}
\eta\partial^2_zv_z=-\frac{A}{2}[z-\frac{2b_1+d}{b_0+b_1+d}\frac{d}{2}].
\end{eqnarray}
In the perfect slip limit (where $A\rightarrow 0$) the
Reynolds condition is automatically fulfilled, i.e., the  solution is self-consistent. However, in general
$\eta\partial^2_zv_z\neq0$ and this limits the validity of the previous results. The Reynolds hypothesis can nevertheless be
justified, for all practical purposes, if
$|\partial^2_zv_z|\ll|\partial^2_zv_i|$, that is, if
$|z-\frac{2b_1+d}{b_0+b_1+d}\frac{d}{2}|\ll|\partial_ig|$. Since
$|\partial_ig|\sim L$, where $L$ is a typical lateral dimension of
the lever, the Reynolds assumption is in general justified for
$L\gg d$.\\
To study the influence of the slip length on the dynamic properties of the oscillator considered in ref.~\cite{nous} we will now consider a simplification.
Such a simplified force expression is obtained if the condition
$b_0=b_1=b$ is fulfilled.
Considering for example the rectangular plate (in the limit given
by Eq.~22) we get for the total force
\begin{eqnarray}
F_z=-\eta ULw\frac{L^2+12bd}{d^3+6bd^2}.
\end{eqnarray}
From this follows the damping constant $\gamma:=-F_z/U$. To
compare this result with the experiment we show on Fig.~2 the
evolution of the quality factor $Q=k/(\omega_0\gamma)$ of the
oscillating beam calculated for a stiffness $k=0.0396$ $N/m$ and a
pulsation $\omega_0=2\pi\times 50$ $kHz$ considered in
ref.~\cite{nous} and for different slip lengths $b$. Clearly, the
agreement with the experimental data is very good in the limit $d
\lesssim \delta_B$ (i.e, far from the observed saturation at large
gaps) for  giant slip lengths such as $b> 500$ $\mu m$ (red
curve). Oppositely the reported results are in total conflict with
the standard no-slip prediction $Q\sim d^3$ (blue curve). While
the present study focussed on the static regime, it is interesting
to remark from dimensional analysis that one should expect to
observe saturation around  $\delta_B\simeq 10$ $\mu$m (i.e.,
$\gamma_{\textrm{lim}}=2\eta S/\delta_B$). The value observed
experimentally corresponds to $\delta_B\simeq 25$ $\mu$m, which is
of the same order of magnitude but nevertheless significantly
larger. The difference could be imputed to geometry
considerations, i.e., to the fact that the dynamics of the fluid
around the lever  should be strongly influenced by the finite size
of the system under study (indeed, in the regime where the
boundary layer $\delta_B$ plays explicitly a role, retardation
should be taken into account). A different explanation could be
that the model of perfect slip breaks down at large gap. A
preliminary analysis in that direction shows that if we conserve
the inertial term in the Navier-Stokes equations we indeed obtain
a saturation regime due to an additional damping occurring on the
length scale $b\sim \delta_B$ and this even if $b$ is infinite in the static
regime.

\section{Discussion}
Fundamentally, the existence of a giant slip length regime is very
surprising and interesting and should therefore be discussed
carefully. Here, we will only review some results which, we think,
are important to justify microscopically the results
discussed in this article and in ref.~\cite{nous}. We remind that from a microscopic point of view, the slippage
coefficient is actually linked to the very nature of the
interaction between the oscillating surface and the air molecules.
Historically, the first theoretical analysis of this phenomenon
goes back to J.~C.~Maxwell and to its kinetic theory of
gases~\cite{Maxwell}. Following this approach one can indeed
distinguish between a specular channel of interaction, for which
the molecules are colliding elastically with the surface, and a
channel of interaction for which molecules are reflected
diffusively from the
wall~\cite{Maxwell,Loeb,Fichman,Arya,Guo,Sharipova,Lauga}. This
second channel is linked to multiple collisions between molecules
and also to adsorption by the surface. The slip length $b$ in this
statistical model is given by the  Maxwell
formula~\cite{Maxwell,Loeb,Fichman,Arya,Guo,Sharipova}:
\begin{equation}
b\simeq\frac{2}{3}\bar{\lambda}\frac{2-p_d}{p_d},
\end{equation}
where $\bar{\lambda}$ is the typical mean free path of gas
molecules (i.e., $\bar{\lambda}\simeq 60$ nm for air in ambient
conditions) and $p_d$ the tangential momentum accommodation
coefficient, i.e., the fraction of those molecules hitting the
surface which are reflected diffusively. Clearly, if $p_d$
vanishes then the slip length is infinite. This suggests that in
the working regime of our mechanical oscillator the molecules are
mainly reflected specularly. Furthermore, recent analysis based on
the fluctuation dissipation theorem and the Green-Kubo
relationship emphasize the importance of several other
microscopical parameters on the molecular dynamics close to a
surface~\cite{Lauga,Charlaix,Neto}. Altogether these studies
provide an estimate for the slip length given by:
\begin{eqnarray}
b\sim\frac{ k_B T\eta D}{C_{\bot}\rho\sigma\epsilon^2},
\end{eqnarray}
where $D$ is the fluid diffusion coefficient, $\sigma,\epsilon$
are respectively a typical length and energy characterizing the
molecular interaction, and $C_{\bot}$ a coefficient measuring the
roughness (larger $C_{\bot}$ mean larger roughness). This shows
the importance of surface roughness but also of surface defects
and nano-structuration~\cite{Charlaix} for the physics of slippage
at solid-gas interfaces. Future work should investigate the
effects of theses  parameters on the damping coefficient $\gamma$.
It is worth noting that past studies on the slippage at a
solid-fluid interface mainly focussed on liquids for which the
mean free path is much smaller ($\bar{\lambda}\sim$ 1 nm) and for
which interactions between molecules are much stronger. The
existence of a partial slip regime implies in those cases to work
in the realm of nanofluidics (e.g., nanochannels) with separating
gap $d$ well below the micrometer range~\cite{Neto,Tabeling} or
with AFMs in contact mode~\cite{Neto,Aime}.
Here oppositely, we consider gases and we can define a Knudsen
coefficient $K_n=\bar{\lambda}/d\sim 0.001-0.06$ which corresponds
to a regime of transition flow \cite{Tabeling,Charlaix,Lauga}
occurring at large gap values $d$ (it is indeed well known  that
important deviations to the no slip boundary conditions appear for
$K_n\sim 10^{-3}-10^{-2}$~\cite{Tabeling}).\\
Another relevant length in the analysis is the vibration amplitude
of the lever which we reported in ref.~\cite{nous} to be $\delta z\simeq 0.05$ nm. This is actually a very small amplitude
which is attainable experimentally mainly because of the thermal excitation mode and high sensitivity of the optical detection setup used in \cite{nous}. This value for $\delta z$ is also comparatively smaller that those attainable with actuated AFM~\cite{Maali}. Working in this regime where $\delta z/\bar{\lambda}\simeq10^{-3}\ll 1$ could therefore lead to new physics and we expect that further studies in this direction will be done in a close future (e.g., to compare the effect of the vibration amplitude and of the excitation modes on the micro- and nano-lever dynamics).\\
In this context it is worth mentioning that the theoretical model developed in this article for the plane-plane configuration  constitutes the equivalent of the Vinogradova formula obtained for the sphere-plane configuration~\cite{Vinogradova,Vinogradova2,Maali,Charlaix,Lauga}. The analysis of Vinogradova  generalizes the result obtained by Taylor~\cite{Neto} in the no slip limit and which predicts a friction force $F_z=-6\pi\eta R^2 U/d$ for a sphere of radius R moving along the $z$ axis with the velocity $U$ perpendicularly to the interface $z=0$ and separated from this plane by a minimal gap $d$. It is interesting to point out that the no-slip condition predicts the same law $F_z\propto 1/d$ as in the plane-plane configuration with the perfect slip condition (i.e., Eq.~28) but with a different numerical value for $\gamma_{\textrm{Taylor}}:=6\pi\eta R^2/d$ (compare Eq.~1). The Vinogradova model predicts oppositely $F_z\simeq 2\pi\eta R^2U\ln{(6b/d)}/b$ in the limit $b\rightarrow + \infty$ of the perfect slip.
\begin{table}
\begin{tabular}{l||c|c}
  & plane-plane & sphere-plane\\
\hline
\hline
$b=0$  & $\gamma=-\eta\langle g\rangle S/d^3$ & $\gamma=6\pi\eta R^2 /d$\\
\hline
$b\rightarrow +\infty$  & $\gamma=2\eta S/d$ & $\gamma=2\pi\eta R^2\ln{(d/(6b))}/b$\\
\end{tabular}
\caption{Table summarizing the asymptotic viscous force regimes for both the plane-plane and sphere-plane configuration and for the no-slip and perfect slip cases. }
\end{table}
The different regimes of force depending of the value for the slippage length $b$ and from the geometry considered are summarized in Tab.~1.\\
The present problem reminds us a very known similar difficulty encountered in experiments for  measuring the Casimir force in the sphere-plane or plane-plane configuration. It is worth pointing out however, that in the Casimir effect~\cite{Genet} the force in the plane-plane configuration varies as $F_z\sim \hbar S/d^4$ whereas it varies as $F_z\sim \hbar R/d^3$ for the sphere-plane configuration ($\hbar$ is the planck constant). Beside the important difference in the power law behavior in $d$ between these expressions and those predicted by hydrodynamics it is interesting to observe that in the perfect slip limit the viscous force decays slowly when $d$ increases whereas the same force vanishes in the sphere-plane as predicted by the Vinogradova formula for $b=+\infty$. Therefore, in order to observe the perfect slip regime the plane-sphere configuration would be much more demanding than the plane-plane configuration studied in this work and in \cite{nous} since it implies that one should consider very small gaps $d$ to obtain a finite effect with the sphere. This could have implication in optical near-field microscopy where viscosity is suggested as a possible mechanism to justify the shear force applied on the tip probes~\cite{Karrai} (see also \cite{Brun1} and \cite{Brun2} for experimental demonstrations in high vacuum at cryogenic temperatures). Additionally, this sensitivity of the force behavior with $b$ in the sphere-plane configuration could be used to probe more precisely the value of the slippage length than in the plane-plane geometry. Oppositely, the giant slip effect studied in this paper could thus be a specificity of the planar geometry and is expected therefore to have a huge impact on the NEMS dynamics which are mostly developed with such geometry.
\section{Implications for NEMS architectures and dynamics}
This brings us to the  second point that we shall now
discuss in this section, that is the implication of our results
for NEMS and MEMS engineering. Owing to the formula given by Eq.~1
the effect of dissipation on the beam motion decays very slowly
with $d$ and in particular we see that the quality factor
$Q:=m\omega_0/\gamma$ decreases linearly with $d$.
\begin{figure}[hbtp]
\includegraphics[width=8.2cm]{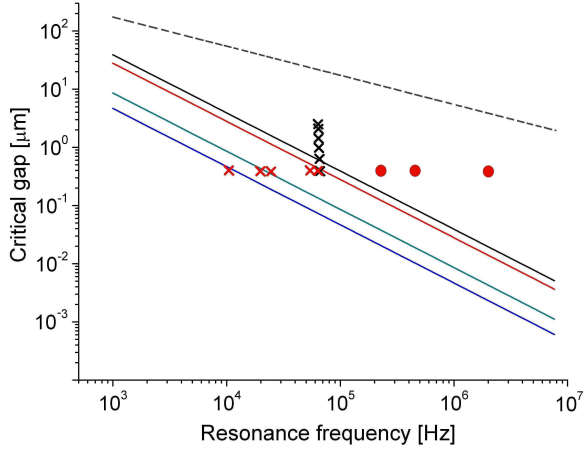}\\
\caption{\label{fig:Dcrit2} Evolution of critical gap as a
function of frequency $f_0=\omega_0/(2\pi)$ for a Si (black line),
SiC (red line), GaAs (green line), and Au (blue line) cantilever,
respectively. The black dashed line corresponds to the limit
associated with the boundary layer. Experimental data points for
the lever studied in ref.~\cite{nous} at the internal resonance
frequency 50 kHz are also shown for distance $d$ close to the
overdamping regime (black crosses). Red crosses and circles
correspond to characteristics of NEMS realized with a distance to
the substrate constantly equal to $d=400 nm$. Such NEMS can
approximately be modeled as the one shown in Fig.~1.}
\end{figure}
Additionally, when $d$ decreases the pulsation at resonance
\begin{equation}
\omega_{\textrm{reson.}}=\sqrt{\omega^2_0-\frac{1}{2}\left(\frac{\gamma}{m}\right)^2}
\end{equation}
is progressively down shifted~\cite{nous}. This occurs until
$\omega_{\textrm{reson.}}=0$, i.e. from Eqs.~1,33 when the critical
distance $d_c$ given by
\begin{equation}
 d_c= \frac{\sqrt{2}\eta Lw}{m\omega_0},
\end{equation} is reached. For $d \leq d_c$ the lever motion
is consequently overdamped (nonlinear effect are also expected in this limit ). To describe quantitatively the
importance of such a regime on beam dynamics we remind that for the
levers considered here we have:
$\omega_0=\sqrt{\frac{E}{12\rho}}r_0^2\cdot \frac{t}{L^2}$ and
$m=\rho t L w$ where $\rho$ is the bulk density, $E$ is the
Young's modulus, and $r_0\simeq 1.875104$~\cite{Cleland}.\\
Fig.~3 shows the variations of $d_c$ as a function of
$f_0=\omega_0/(2\pi)$ for a thin lever with $t=180$ nm (same as in
ref.~\cite{nous}) and for different commonly used materials. For
low frequencies in the 10 kHz range and below, the overdamping
regime appears already at large separation distance $d_c\geq$
0.1-1 $\mu$m. Oppositely, for very high oscillator frequencies in
the 100 kHz range and beyond we have  $d_c \lesssim 10-100$ nm and
the overdamping regime becomes a fundamental issue only at the
nanoscale. For comparison we show on the same graph the
experimental data points taken from ref.~\cite{nous} and
corresponding to working distances $d$ which are decreasing until
the overdamped regime at $d_c$ is reached. Additionally, we show
also the physical characteristics recorded (i.e., internal
resonance frequency $f_0$, and fixed distance gap $d=400$ nm with
the substrate) of typical Si made NEMS. These NEMS can be with a
good approximation described with the simple geometry considered
here. Clearly, working with such NEMS in a gaseous environment may
strongly affect their dynamics and only for very high $f_0$ could
the overdamping regime actually be overcome. This is indeed
confirmed for those NEMS annexed by a red cross in Fig.~3 which we
studied experimentally by using the same optical method as
described in ref.\cite{nous}. The experiment showed that NEMS with
such gaps do not resonate in  air at room temperature confirming
therefore the role played by overdamping. For the NEMS indicated
by a red circle, \emph{i.e.}, with  frequency in the 100 kHz and
MHz range we were out of the detection sensitivity of our setup
and no data were available.\\
\begin{figure}[hbtp]
\includegraphics[width=1\columnwidth]{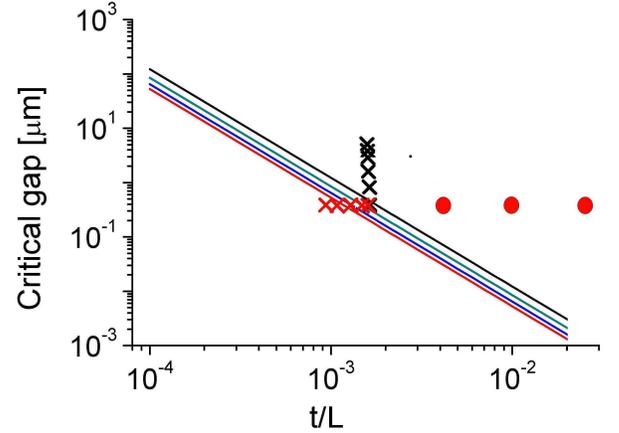}\\
\caption{\label{fig:Dcrit2} Evolution of the critical gap as a
function of the ratio $t/L$ for the same lever materials as in
Fig.~2. Colors lines and data points have the same meaning as in
Fig.~2.}
\end{figure}
As a complementary analysis we show in Fig.~4 the explicit
dependence of $d_c$ on the aspect ratio $t/L$ for different
materials. In the typical range of aspect ratio considered the
overdamping regime covers distance gaps $d$ going from the
micrometer range to the nanoscale and, therefore, cannot be
neglected. Again, this fact is confirmed by comparing these graphs
with
available experimental data (see Fig.~4 and compare with Fig.~3).\\
Furthermore, it is also useful to remind once again that for a large gap
$d$ the important length scale is the boundary layer thickness
$\delta_B$ which characterizes the spatial region surrounding the
lever for which viscosity has an impact on the fluid
dynamics~\cite{landau,batchelor}. For $d\geq \delta_B$ the
substratum lies outside this layer and dissipation must
saturate~\cite{landau,batchelor}, as reported in ref.~\cite{nous}.
Comparisons with values for $d_c$ (see Fig.~3) show that the
overdamping regime is always reached for gaps smaller than
$\delta_B$. The overdamping regime appears consequently as
a robust limitation which should affect the design of any NEMS
working in fluids. The results obtained here for a particular
lever geometry are expected to be very general as soon as the beam
geometrical dimensions are larger than $\delta_B$.
However, when dimensions are smaller, boundary effects due to the
finite size of the system should explicitly be taken into account
in the definition of $d_c$.
\section{Conclusion}
In this article we studied the linearized Navier-Stokes equation in the static regime to describe the damping mechanism of oscillating micro plate in air  close to a substrate. We considered the influence of the slip length and showed that results reported in ref.~\cite{nous} are only compatible with very large slip length in the range $b> 500\mu$m. We discussed the implication of this mechanism on the oscillation properties of NEMS and showed that an overdamping behavior represents a fundamental mechanism for such systems. We expect that this work could have important consequences for NEMS engineering.
\begin{acknowledgments}
This research was partly supported by a ``Carnot-NEMS''
collaborative grant between CEA-LETI and Institut N\'eel.
\end{acknowledgments}
\appendix
\section{The rectangular plate in the static regime}
The aim of this appendix section is to solve the Poisson equation $(\partial_x^2+\partial_y^2)p(x,y)=A$ where $A$ is a constant for the rectangular domain $L\times w$ with boundary conditions $p=p_0$ along the contour.
Using a Fourier series satisfying these boundary conditions at $x=0$ and $x=L$ we can write
\begin{eqnarray}
p(x,y)=p_0+\sum_n c_n(y)\sin{(\frac{\pi n x}{L})}.
\end{eqnarray}
Here $c_n(y)$ is solution of
\begin{eqnarray}
-(\frac{\pi n }{L})^2 c_n(y)+ \frac{d^2}{dy^2}c_n(y)=\frac{4A}{n\pi}\epsilon_n
\end{eqnarray}
with $\epsilon_n=1$ if $n$ is a odd integer and $\epsilon_n=0$ if $n$ is even.
The general solution of this equation is
\begin{eqnarray}
c_n(y)=a_n^{(+)}e^{+\pi n y/L}+a_n^{(-)}e^{-\pi n y/L}-\frac{4A}{n^3\pi^3}L^2\epsilon_n.
\end{eqnarray}
The constant $a_n^{(\pm)}$ are determined by the boundary
conditions $p=p_0$ at $y=0$ and $y=w$. One therefore obtain
\begin{eqnarray}
a_n^{(-)}=\frac{4A}{n^3\pi^3}L^2\epsilon_n\frac{[e^{+\pi n w/L}-1]}{[e^{+\pi n w/L}-e^{-\pi n w/L}]}\nonumber\\
a_n^{(+)}=\frac{4A}{n^3\pi^3}L^2\epsilon_n\frac{[1-e^{-\pi n w/L}]}{[e^{+\pi n w/L}-e^{-\pi n w/L}]}.
\end{eqnarray}
This leads to
\begin{eqnarray}
p(x,y)=p_0+\frac{4AL^2}{\pi^3}\sum_n\frac{\epsilon_n}{n^3}\sin{(\pi
nx/L)}[-1\nonumber\\
+\frac{1-e^{-\pi n w/L}}{e^{+\pi n w/L}-e^{-\pi
nw/L}}e^{+\pi ny/L}\nonumber\\+\frac{e^{+\pi n w/L}-1}{e^{+\pi n
w/L}-e^{-\pi n w/L}}e^{-\pi n y/L}].
\end{eqnarray}


\begin{thebibliography}{}
\bibitem{Roukes2007}
M.~Li, H.~X.~Tang, and M.~L.~Roukes, Nature Nanotech.~{\bf 2}, 114 (2007).

\bibitem{cooling} C.~Metzger, \emph{et al.}, Phys.~Rev.~Lett.~{\bf 101}, 133903 (2008); G.~Jourdan, F.~Comin, and J.~Chevrier, Phys. Rev. Lett. {\bf 101}, 133904 (2008)

\bibitem{Ekinci}
Y.~T.~Yang, \emph{et al.}, Nano Lett.~{\bf 6}, 583 (2006).

\bibitem{Cap1}
K.~L.~Ekinci,Y.~T.~Yang ,and M.~L.~Roukes, J.~Appl.~Phys {\bf 95}, 2682 (2004).

\bibitem{Cap3}
J.~N.~Munday, F.~Capasso, and A.~Parsegian, Nature (London) {\bf 457}, 170 (2009).

\bibitem{rugar01}
H.~J.~Mamina and D.~Rugar, Appl.~Phys.~Lett. {\bf  79}, 3358 (2001).

\bibitem{Tabeling}
P.~Tabeling, \emph{Introduction to microfluidics} (Oxford
University Press, USA, 2006).

\bibitem{nanoc}
P.~Joseph, \emph{et al.}, Phys.~Rev.~Lett. \textbf{97}, 156104 (2006).

\bibitem{Lauga}
E.~Lauga in \emph{Handbook of Experimental Fluid dynamics} edited by J.Foss, C.Tropea and A.Yarin, (Springer, New York 2007), Chapter 19 pp.~1219.

\bibitem{Green}
C.~P.~Green and J.~E.~Sader, J.~Appl.~Phys. {\bf 98}, 114913 (2005).

\bibitem{Naik}
T.~Naik, E.~K.~Longmire, and S.~C.~Mantell, Sensor and Actuators A: Physical {\bf 102}, 240 (2003).

\bibitem{Paul}
M.~R.~Paul and M.~C.~Cross, Phys.~Rev.~Lett. {\bf 92}, 235501 (2004).

\bibitem{Dorignac}
J.~Dorignac, \emph{et al.}, Phys.~Rev.~Lett. {\bf 96}, 186105 (2006).

\bibitem{Basak}
S.~Basak, A.~Raman, and S.~V.~Garimella, J.~Appl.~Phys. {\bf 99}, 114906 (2006).

\bibitem{Tung}
R.~C.~Tung, A.~Jana, and A.~Raman,  J.~Appl.~Phys. {\bf 104}, 114905 (2008).

\bibitem{nous}
A.~Siria, \emph{et al.}, Phys.~Rev.~Lett.~\textbf{102}, 254503 (2009).

\bibitem{Vinogradova}
O.~I.~Vinogradova, Langmuir \textbf{11}, 2213 (1995).
\bibitem{Vinogradova2}
O.~I.~Vinogradova, Langmuir \textbf{14}, 2827 (1998).

\bibitem{landau} L.~D.~Landau and E.~M.~Lifshitz,
\emph{Fluid Mechanics} (Pergamon, Oxford, 1975).

\bibitem{batchelor}
G.~K.~Batchelor, \emph{Fluid dynamics}, (Cambridge University
Press, Cambridge, UK, 1974).

\bibitem{Loeb}
L.~B.~Loeb,\emph{The Kinetic Theory of Gases}, (McGraw-Hill, New York, USA, 1927).

\bibitem{Neto}
C.~Neto \emph{et al.}, Rep.~Prog.~Phys.~\textbf{68}, 2859 (2005).

\bibitem{Charlaix}
L.~Bocquet, E.~Charlaix,  to appear in Chem.~Soc.~Rev. (2010); DOI:10.1039/B909366B.

\bibitem{Navier}
C.~L.~M.~H.~Navier, M\'{e}moire de l'Acad\'{e}mie Royale des Sciences de l'Institut de France \textbf{6}, 389-440 (1823).
\bibitem{remarque}
The pressure can be deduced from the equation $\nabla^2 p=0$ which
results from the Navier-Stokes equation~\cite{landau,batchelor}.
Supposing $\partial_z p=0$  leads to
$(\partial_x^2+\partial_y^2)p=0$. Applying the boundary condition
$p=p_0$ along the contour together with the uniqueness theorem
implies directly $p=p_0$ everywhere. Finally, it is worth noting
that the solutions of the Navier-Stokes equations
$\partial_z^2v_i=0$ $(i=x,y)$ are
$v_i(x,y,z)=\alpha_i(x,y)z+\beta_i(x,y)$. The perfect slip
boundary conditions then leads to $\alpha_i=0$, and the
imcompressibility relation to
$\partial_x\beta_x+\partial_y\beta_y=-U/d$.
\bibitem{Maxwell}
J.~C.~Maxwell, Phil.~Trans.~R.~Soc.~Lond. \textbf{170}, 231 (1879).
\bibitem{Fichman}
M.~Fichman and G.~Hetsroni, Phys.~Fluids \textbf{17}, 123102 (2005).
\bibitem{Arya}
G.~Arya \emph{et al.}, Molecular Simulation \textbf{29}, 697 (2003).
\bibitem{Guo}
Z.~L.~Guo \emph{et al.}, Europhys.~Lett.~\textbf{80}, 24001 (2007).
\bibitem{Sharipova}
F.~Sharipova and D.~Kalempa, Phys.~Fluids \textbf{15}, 1800 (2003).
\bibitem{Aime}
A.~Maali, \emph{et al.}, Phys.~Rev.~Lett.~{\bf 96}, 086105 (2006).
\bibitem{Maali}
A.~Maali and B.~Bhushan, Phys.~Rev.~E \textbf{78}, 027302 (2008).
\bibitem{Genet}
C.~Genet \emph{et al.}, Annales Fondation L. de Broglie, \textbf{29}, 331 (2004).
\bibitem{Karrai}
K.~Karrai and I.~Tiemann, Phys.~Rev.~B \textbf{62}, 13174 (2000).
\bibitem{Brun1}
M.~Brun, \emph{et al.}, J. Microscopy \textbf{202}, 202 (2001).
\bibitem{Brun2}
M.~Brun, \emph{et al.}, Europhys.~Lett. \textbf{64}, 634 (2003).
\bibitem{Cleland}
A.~N.~Cleland, \emph{Foundation of nanomechanics}
(Springer-Verlag, Germany, 2004).
\end{thebibliography}
\end{document}